\documentclass[prl, reprint, 10pt, english,british]{revtex4-1}
\pdfoutput=1	
\usepackage[T1]{fontenc}
\usepackage[latin9]{inputenc}
\usepackage[letterpaper]{geometry}
\usepackage{amssymb}
\usepackage{amsbsy}
\usepackage{graphicx}
\usepackage{setspace}
\usepackage{esint}

\makeatletter

\providecommand{\tabularnewline}{\\}

\makeatother

\usepackage{babel}
\begin{document}

%
%
%

\title{Complete photoionization experiments via ultra-fast coherent control with polarization-multiplexing}

\author{P. Hockett}
\email[Email: ]{paul.hockett@nrc.ca}
\affiliation{National Research Council of Canada, 100 Sussex Drive, Ottawa, K1A 0R6, Canada}
\author{M. Wollenhaupt}
\affiliation{Institut für Physik, Carl von Ossietzky Universität Oldenburg, Carl-von-Ossietzky-Straße 9-11, 26129 Oldenburg, Germany}
\author{C. Lux}
\author{T. Baumert}
\affiliation{Institut für Physik, Universität Kassel, Heinrich-Plett-Str. 40, 34132
Kassel, Germany}

\date{23rd Feb. 2014}

\begin{abstract}
Photoelectron angular distributions (PADs) obtained from ionization
of potassium atoms using moderately intense femtosecond IR fields ($\sim$10$^{12}$Wcm$^{-2}$)
of various polarization states are shown to provide a route to ``complete''
photoionization experiments. Ionization occurs by a net 3-photon absorption process, driven via the $4s\rightarrow4p$ resonance at the 1-photon level. A theoretical treatment incorporating the intra-pulse electronic dynamics allows for a full set of ionization matrix elements to be extracted from 2D imaging data. 3D PADs generated from the extracted matrix elements are also compared to experimental, tomographically reconstructed, 3D photoelectron distributions, providing a sensitive
test of their validity. Finally, application of the determined matrix
elements to ionization via more complex, polarization-shaped, pulses
is demonstrated, illustrating the utility of this methodology towards
detailed understanding of complex ionization control schemes and suggesting
the utility of such ``multiplexed'' intra-pulse processes as powerful
tools for measurement.
\end{abstract}


\maketitle

So-called ``complete'' measurements of ionization dynamics aim
to obtain the amplitudes and phases of the ionization matrix elements
which describe the ionization event in terms of the partial wave decomposition
of the outgoing photoelectron \cite{Reid2003,Cherepkov2005}. Since
determining the phases requires an observable in which interferences
between partial waves are present, photoelectron angular distributions
(PADs) are required for complete measurements. Such experiments have
long been performed for atomic systems \cite{Becker1998,Reid2003}
where, for example, PADs obtained via pump-probe schemes utilising
linearly polarized light and a range of pump-probe geometries \cite{berry1976,Hansen1980,Chien1983},
or different polarization states \cite{Duong1978}, have allowed the
relative amplitude and phase of two ionization matrix elements to
be determined \cite{Lambropoulos1973}. For molecules, various experimental
techniques, including molecular frame PADs \cite{Gessner2002,Lebech2003,Yagishita2005},
time-resolved rotational wavepacket studies \cite{Suzuki2006,Suzuki2012}
and state-resolved measurements \cite{reid1991,Reid1992,Hockett2009},
have been demonstrated. The common thread to all of these measurements
is the necessity of a data-set containing sufficient information to
reliably obtain the set of ionization matrix elements (which may be
large) via some type of fitting procedure. Here ``sufficient''
refers to both the size of the experimental data-set and the fundamental level
of detail present \cite{Cherepkov2005}. For example, measurements
from the Zare group demonstrated the level of detail obtainable from experiments with linear and elliptically polarized light \cite{reid1991,Reid1992}; work from Elliott's group investigated the role of interferences between 1 and 2 photon ionization pathways, including control of PADs \cite{Yin1995, Wang2001}.

Control experiments have garnered much interest over the last 20 years,
in particular with the aim of controlling the outcome of chemical
reactions, as well as control over observables such as PADs \cite{Yin1992,Yin1995,Wollenhaupt2009a,Krug2009,Wollenhaupt2013}.
Control experiments utilizing shaped laser pulses can also be considered
as intra-pulse pump-probe experiments; in this context, such experiments
may be highly-multiplexed, and could be considered as a natural continuation
of (serial) frequency-domain schemes for measuring atomic and molecular
properties. Furthermore, measurement of the relevant physical properties
allows the control processes to be understood in detail, rather than
treated as a black-box optimization scheme, as pointed out in refs. \cite{Wollenhaupt2005b,Wollenhaupt2011}.
In this work we demonstrate this principle using an experimental scheme
originally designed with the aim of controlling PADs \cite{Wollenhaupt2009a,Wollenhaupt2013}
but, instead, make use of the data - measured as a function of polarization
state - to elucidate the ionization matrix elements. Because of
the high information content of the PADs obtained, four experimental
measurements prove sufficient for a determination of the ionization
matrix elements. The treatment can readily be extended to more complex,
arbitrarily shaped pulses, and also to molecular ionization in cases
where the intra-pulse dynamics is computationally tractable, allowing
for truly multiplexed measurements beyond the proof-of-concept shown
here. We demonstrate this principle by application of the determined
matrix elements to ionization via a polarization-multiplexed pulse.

The experimental set-up has been covered in detail in refs. \cite{Wollenhaupt2009a,Wollenhaupt2009b,Krug2009}.
Here we briefly outline the control scheme for the case of polarization
shaped pulses. Moderately intense ($\sim$10$^{12}$~Wcm$^{-2}$)
laser pulses (795~nm, 30~fs, bandwidth $~$60~meV FWHM) were generated by a
Ti:Sapphire multi-pass amplifier. 
Basic control of the ellipticity of the pulse was achieved via a $\lambda/4$
plate, while more complex pulse shapes were attained with the use
of a spatial light modulator in a 4\emph{f} configuration \cite{Wollenhaupt2009a,Praekelt2003}.
The laser pulses were focussed into potassium vapour generated by a
dispenser source in the interaction region of a velocity-map imaging
spectrometer. The detector consisted of a dual-MCP stack, phosphor
screen and CCD camera, allowing measurement of 2D projections of the
full 3D photoelectron distribution. In the case of cylindrically symmetric
distributions, a single 2D projection is sufficient to reconstruct
the full 3D distribution via standard inversion techniques \cite{Whitaker2003};
for non-cylindrically symmetric distributions projections at several
different angles to the detector must be obtained, and a tomographic
reconstruction technique applied to obtain the original 3D distribution
\cite{Wollenhaupt2009b,Smeenk2009,Hockett2010}.

\begin{figure}
\begin{centering}
\includegraphics[scale=0.8]{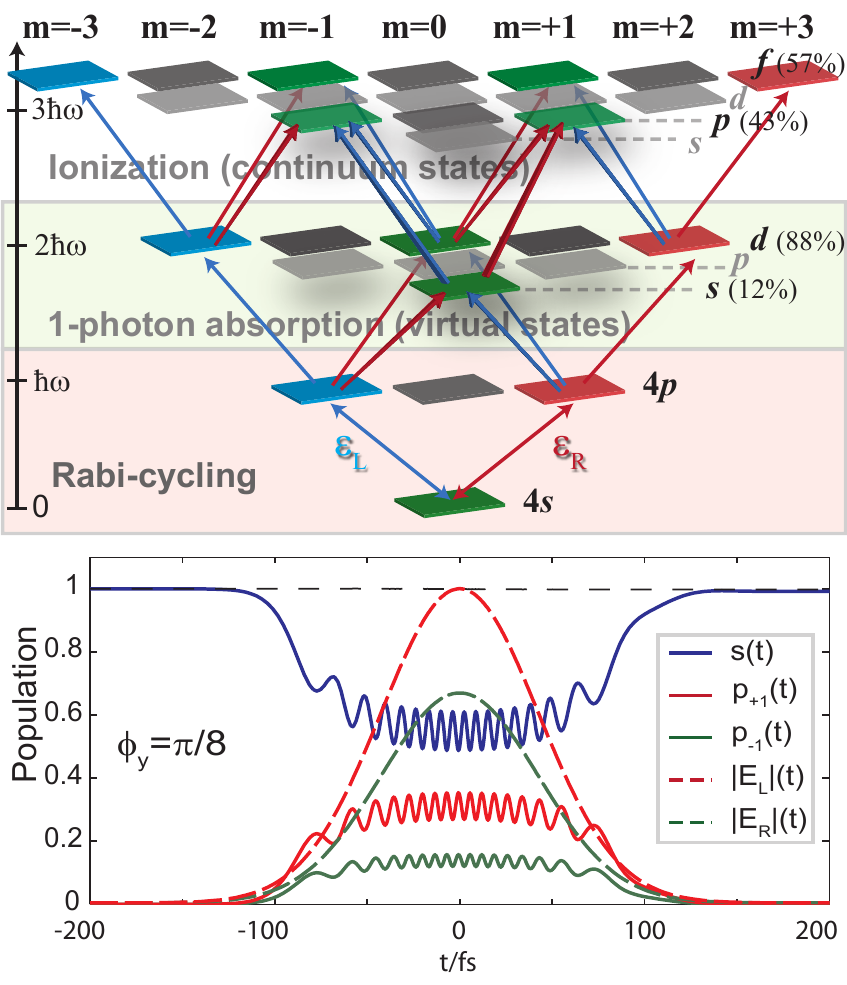}
\par\end{centering}

\caption{Schematic of the 3-photon ionization scheme for potassium. $|l,m\rangle$
states coloured red or blue are accessed by $E_{R}$ or $E_{L}$
components of the field only, while states coloured green can be accessed
by both components via different pathways. States coloured
dark grey are inaccessible $m$ levels within a populated $l$ manifold,
while light grey states show inaccessible $l$ manifolds.  Percentages
give the $m$-summed weightings of the $l$ states as determined from
the fitted radial matrix elements, as listed in table \ref{tab:Fitted-params};
the continuum state populations are also summed over all paths, see
main text for details. The lower panel shows an example of the population
dynamics, in this case for an elliptically polarized pulse $\phi_{y}=\pi/8$, where $\phi_{y}$ is the spectral phase applied to the $y$-component of the $E$-field and is used to define polarization states throughout this work \cite{Wollenhaupt2009a};
the envelope of the laser pulse is also
shown in terms of the $E_{L}(t)$ and $E_{R}(t)$ components. Due to the detuning of the frequency of the laser pulse with respect to the atomic transition the time evolution of the population is characterized by coherent population return rather than Rabi oscillations.
\label{fig:ionization-paths}}

\end{figure}

In order to understand and treat the \emph{intra}-pulse light-matter interaction
(i.e. simultaneous excitation and ionization dynamics) we split the problem conceptually
into two steps, (1) a non-perturbative absorption at the 1-photon
level, the ``pump'' step, (2) a perturbative 2-photon ionization,
the ``probe'' step \cite{Wollenhaupt2003}. This is essentially an \emph{intra}-pulse 1+2 REMPI (resonance-enhanced multi-photon ionization) scheme, where the first step is near-resonant with the potassium $4s\rightarrow4p$ transition (which carries significant oscillator strength), and
the second step is non-resonant. A schematic of the ionization
pathways for this net 3-photon absorption process is given in figure \ref{fig:ionization-paths}(upper
panel).

The ``pump'' process describes the interaction of the control field
with the atom at the 1-photon level. With a moderately intense, near
resonant field, Rabi oscillations are driven. These oscillations follow
the driving electric field and, crucially, depend sensitively on the
instantaneous properties of the light field, such as the polarization
state and the frequency spread. The population dynamics during the
laser pulse are then given by the time-dependent Schrödinger equation:

\begin{equation}
\frac{d}{dt}\left(\begin{array}{c}
s(t)\\
p_{+1}(t)\\
p_{-1}(t)
\end{array}\right)=i\left(\begin{array}{ccc}
0 & \frac{1}{2}\Omega_{L}^{*}(t) & \frac{1}{2}\Omega_{R}^{*}(t)\\
\frac{1}{2}\Omega_{L}(t) & \delta_{+1} & 0\\
\frac{1}{2}\Omega_{R}(t) & 0 & \delta_{-1}
\end{array}\right)\left(\begin{array}{c}
s(t)\\
p_{+1}(t)\\
p_{-1}(t)
\end{array}\right)\label{eq:TDSE}
\end{equation}

where $s(t)$, $p_{+1}(t)$ and $p_{-1}(t)$ are the state vector
components for the 4$s$ and 4$p(m=\pm1)$ states; $\Omega_{L/R}(t)=\mu_{L/R}E_{L/R}(t)$
are Rabi frequencies, where $\mu_{L/R}$ are the transition amplitudes,
and $E_{L/R}(t)$ represents the electric field expanded in a spherical
basis; $\delta_{\pm1}$ represents the detuning of the laser from
the resonant frequency of the transition. In this work $\mu_{L/R}$,
$E_{0}$ (total electric field strength) and $\hbar$ are all set
to unity. For determination of PADs these simplifications are acceptable
as only the \emph{relative} population of $m=\pm1$ states will affect
the angular distribution 
\footnote{For the same reason the value of $\delta_{\pm1}$ has no effect on the PADs in this case, except as a scaling factor for the total population transfer at the 1-photon level; in the calculations shown in figure \ref{fig:ionization-paths} $\delta_{\pm1}=0.05$~rad/fs. For completeness, we also note that spin is neglected in this treatment.}. In this case the relative populations are dependent only on the driving
laser field polarization; in a case where additional $|l,m\rangle$ states
are accessed, more careful treatment of the transition amplitudes
$\mu_{L/R}$ would be required. An example of the population dynamics
is given in figure \ref{fig:ionization-paths}(lower panel), 
 illustrating how the difference
in magnitudes of the $E_{L}(t)$ and $E_{R}(t)$ laser field components
describing an elliptically polarized field give rise to different
$p_{+1}$ and $p_{-1}$ populations.

The ``probe'' process describes the subsequent absorption of 2 photons
resulting in ionization of the 4$p$ excited state. By treating this
step pertubatively the resultant matrix elements are independent of
the instantaneous pulse intensity, hence are assumed to be constant
over the pulse envelope. In the perturbative regime, the 2-photon
dipole transition amplitude for a transition from state $|l_{i},m_{i}\rangle$
to a final state $|l_{f},m_{f}\rangle$, via a virtual intermediate
state $|l_{v},m_{v}\rangle$, integrated over the pulse duration,
can be written as:

\begin{widetext}
\begin{equation}
d_{l_{f}m_{f}}(k)=\int d_{i\rightarrow v}(k,t)d_{v\rightarrow f}(k,t)dt=\int dt\sum_{\begin{array}{c}
l_{i},m_{i};
l_{v},m_{v}\\
q,q'
\end{array}}R_{l_{v}l_{f}}(k)\langle l_{f}m_{f},1q'|l_{v}m_{v}\rangle R_{l_{i}l_{v}}(k)\langle l_{v}m_{v},1q|l_{i}m_{i}\rangle E_{q'}(t)E_{q}(t)p_{m_{i}}(t)\label{eq:dtInt}
\end{equation}
\end{widetext}

where the summation is over all pathways from the initial, ionizable,
states $|l_{i},m_{i}\rangle$, weighted by their populations $p_{m_{i}}(t)$,
and all polarization states $q$. Here $R_{l_{i}l_{f}}(k)$ are the
radial components and $\langle l_{f}m_{f},1q|l_{i}m_{i}\rangle$ are
Clebsch-Gordan coefficients which describe the angular momentum coupling;
the photoelectron kinetic energy dependence on time, given as $e^{i\delta\omega_{e}t}$
in ref. \cite{Wollenhaupt2009a}, has been omitted. This term is identical
for all $|l,m\rangle$ channels, so will not affect the PADs in the
case of polarization shaped pulses. We have assumed that the angular
part of both bound-virtual and virtual-free transitions can be be
described by matrix elements of the same form; we furthermore assume
a single active electron picture with no angular momentum coupling
of the virtual and final one-electron states to the nascent ion core.
 In general both of these simplifications could be removed, resulting
in a more complex angular momentum coupling scheme and, possibly,
more partial wave components due to additional electron-ion scattering.

The observed photoelectron yield as a function of angle, for a single
$k$ or small energy range $dk$ over which we assume the $R(k)$
can be regarded as constant, is then given by the coherent square
over all final (photoelectron) states:

\begin{eqnarray}
I(\theta,\phi;\, k) & = & \sum_{\begin{array}{c}
l_{f},m_{f}\\
l_{f}^{'},m_{f}^{'}
\end{array}}d_{l_{f}m_{f}}(k)Y_{l_{f}m_{f}}(\theta,\phi)d_{l_{f}^{'}m_{f}^{'}}^{*}(k)Y_{l_{f}^{'}m_{f}^{'}}^{*}(\theta,\phi)\label{eq:Itp}
\end{eqnarray}
where the $Y_{lm}(\theta,\phi)$ are spherical harmonics which describe
the angular form of the photoelectron partial waves.

Finally, we note that the PAD can also be described phenomenologically
by $\beta_{LM}$ parameters \cite{Reid2003}, where:

\begin{equation}
I(\theta,\phi;\, k)=\sum_{L,M}\beta_{LM}(k)Y_{LM}(\theta,\,\phi)\label{eq:IBlm}
\end{equation}
The information content of the observed PADs can be considered in
terms of the number of $L,\, M$ terms present in this expansion,
which will therefore depend on both the inherent properties of the
system and the laser pulse parameters, as shown in equation \ref{eq:dtInt}.
For example, use of linearly polarized light restricts eqn. \ref{eq:IBlm}
to terms with $M=0$ only.

In the preceding treatment the angular momentum coupling is calculated
analytically, and the population dynamics numerically; in both cases
these computations are somewhat routine and are expected to be accurate.
Therefore, only the $R_{ll}$ remain as unknowns: determination of
these complex radial matrix elements is the aim of ``complete''
photoionization experiments. To generate distributions to compare
with the experimental results, PADs calculated according to equation
\ref{eq:Itp} were convoluted with a Gaussian radial distribution
(defined in energy space) to generate gridded volumetric data; 2D
image-plane projections were generated by summation of the 3D distributions.
Although numerically intensive, this procedure generates 2D projections
that can be compared directly with the 2D experimental imaging data
even in the case of non-cylindrically symmetric distributions. In
order to determine the $R_{ll}$, fitting to the experimental data
was carried out 
to optimize the computed 2D projections. In this procedure the $R_{ll}$ were
expressed in magnitude and phase form, $R_{ll}=|R_{ll}|e^{i\delta_{ll}}$,
where $-\pi\leq\delta_{ll}\leq\pi$; also $R_{l_{1}l_{2}}=R_{l_{2}l_{1}}^{*}$.
Because absolute phases cannot be determined, $\delta_{01}$ at the
1-photon level was set to zero as a reference phase. Technical details of this procedure will be given in a future publication \cite{hockett2014}.

\begin{figure*}  
\begin{centering}
\includegraphics{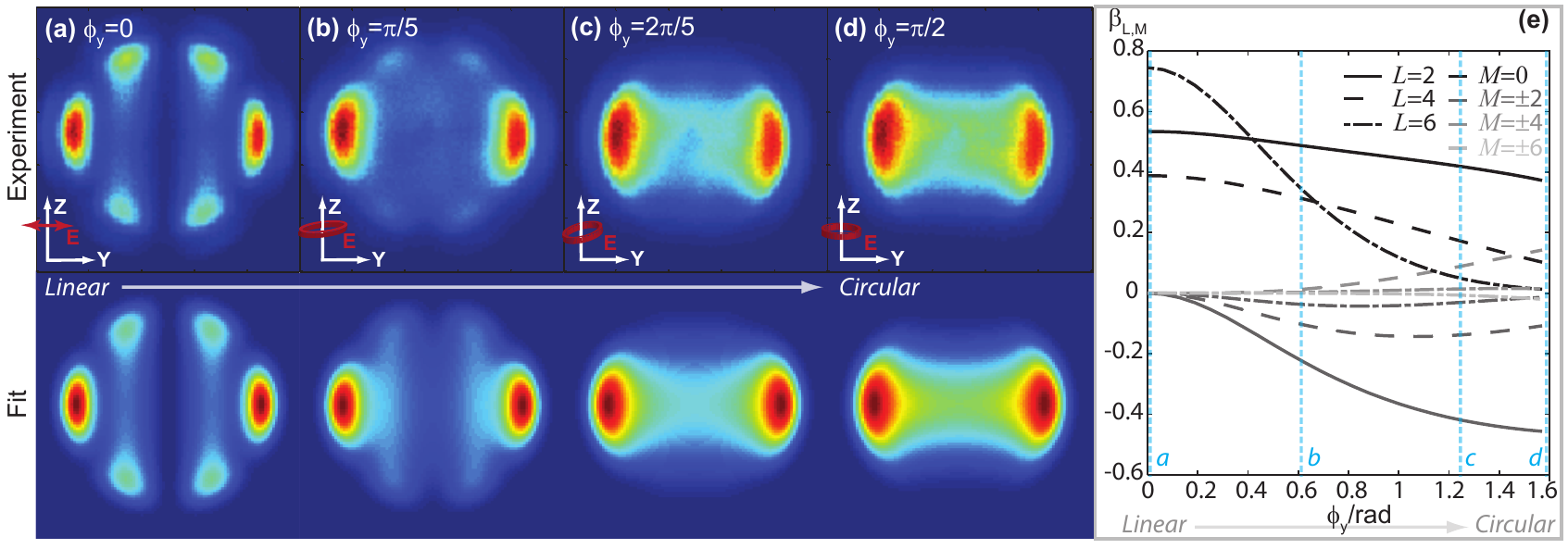}
\par\end{centering}

\caption{(a) - (d) raw data and fit result for the four polarizations investigated.
The laser propagates along the $z$-axis and the polarization state,
defined in the $(x,y)$ plane, is illustrated schematically as the
polarization varies from linear (a) to circular (d); electrons are
detected/integrated in the $(y,z)$ plane. (e) Calculated $\beta_{LM}$
parameters as a function of polarization ($\phi_{y}$), line styles
denote $L$ and shading $M$. The polar axis for the $\beta_{LM}$
is chosen to match the $y$-axis in (a). \label{fig:images}}
\end{figure*}

The best fit images are shown in figure \ref{fig:images}(a)-(d).
The fitted results show a reasonable agreement with the experimental
data in terms of the form of the angular distributions, and trend
with polarization, but less satisfactory agreement in terms of the
width and scaling of the features. For the elliptically polarized
cases this is not too surprising, as the polarization states used
are only approximate. The linear and circularly polarized cases show
a better, but still imperfect, agreement with the data; this is attributed
to the assumption of a Gaussian radial distribution. Nonetheless,
the obtained ionization matrix elements appear to be relatively insensitive
to these issues because they are primarily defined by the angular
coordinate of the image. The calculated $\beta_{LM}$, as a function
of $\phi_{y}$, are shown in figure \ref{fig:images}(e). 
Non-zero values are found for $L=0,\,2,\,4,\,6$ and even $M$ terms, consistent
with the experimental symmetry and the total number of photons absorbed
\cite{Reid2003}, and a smooth variation in the parameters is observed
with ellipticity.

The fitted parameters obtained are given in table \ref{tab:Fitted-params}.
Here the magnitudes are normalized to give total cross-sections of
unity at the 1-photon and 2-photon level. The results show that the
final photoelectron wavefunction, summed over all paths to each final
state $|R_{l_{i}l_{f}}|^{2}=(\sum_{v}|R_{l_{i}l_{v}}||R_{l_{v}l_{f}}|)^{2}$,
is 57\% $f$-wave and 43\% $p$-wave in character. This is consistent
with the expectation from the shape of the PAD, which has strong $L=6$
character, that the $f$-wave dominates, but also reveals a significant
contribution from $l=1$ partial waves, primarily via the $p\rightarrow s\rightarrow p$
channel. The relative phases of the final continuum waves are quite
different, revealing partially destructive interference between the
$p$ and $f$-waves. It is interesting to note that since the scattering
phase and Wigner delay are directly related \cite{wigner55}, these
phase differences indicate a significant difference in emission time
of all the continuum waves, including a dependence on the virtual
state since the $p\rightarrow s\rightarrow p$ and $p\rightarrow d\rightarrow p$
paths accumulate different total phases %
\footnote{We note for completeness that because the Wigner delay is the energy-derivative
of the scattering phase it is not defined in absolute terms by the relative phases at a single-energy, but relative emission times may be determined from a set of such results.}.

\begin{table}
\begin{centering}
\begin{tabular}{c|c|c|c|c|c}
\multicolumn{3}{c|}{Transition} & $|R_{l_{1}l_{2}}|$ & $|R_{l_{1}l_{2}}|^{2}$/\% & $\delta_{l_{1}l_{2}}$/rad.\tabularnewline
\multicolumn{1}{c}{} & \multicolumn{1}{c}{$l_{1}$} & $l_{2}$ &  &  & \tabularnewline
\hline 
\hline 
$i\rightarrow v$ & p & s & 0.34 (3) & 12 (4) & 0 {*}\tabularnewline
\hline 
 & p & d & 0.94 (8) & 88 (11) & -1.62 (4)\tabularnewline
\hline 
$v\rightarrow f$ & s & p & 0.85 (8) & 72 (12) & -0.19 (3)\tabularnewline
\hline 
 & d & p & 0.14 (2) & 2 (2) & -2.08 (8)\tabularnewline
\hline 
 & d & f & 0.51 (9) & 26 (13) & 0.24 (7)\tabularnewline
\end{tabular}
\par\end{centering}

\caption{Fitted values for the relative transition matrix element magnitudes,
$|R_{ll}|$, and phases, $\delta_{ll}$. The square of the magnitudes
is expressed as a percentage of the total transition amplitude, normalized
to unity for each step (these percentages are also shown in figure
\ref{fig:ionization-paths}). Uncertainties in the last digit are
given in parentheses. {*} reference phase, set to zero during fitting.\label{tab:Fitted-params}
}
\end{table}

\begin{figure}
\begin{centering}
\includegraphics[scale=0.7]{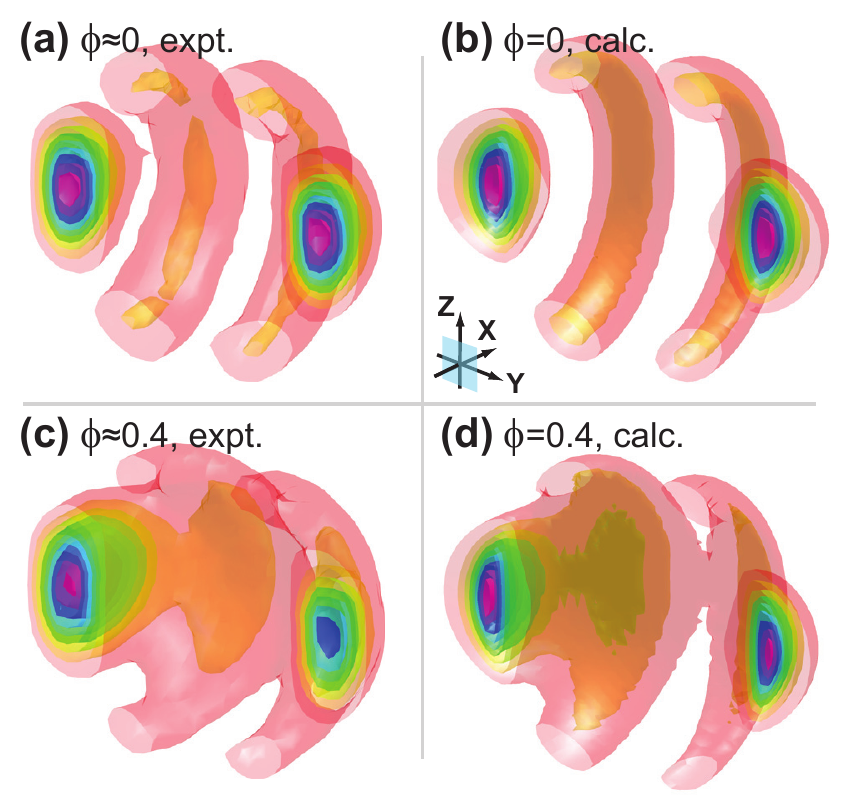}
\par\end{centering}

\caption{
Tomographically reconstructed data \& calculated distributions based
on the fitted ionization matrix elements. The 3D distributions are sliced in order to reveal the radial distribution
in detail; the nested isosurfaces run from 10\% (pale red) to 90\%
(purple) density. \label{fig:tomo-data}
}
\end{figure}

To further validate these results we next consider in detail the
full 3D distributions. Figure \ref{fig:tomo-data} shows the 3D photoelectron
distributions for two polarization states. The experimental data,
fig. \ref{fig:tomo-data}(a) \& (c), was tomographically reconstructed
from a set of 2D images, as detailed in ref. \cite{Wollenhaupt2009b}.
The tomographic data provides a full 3D map of the PADs, providing
details which may be obscured in single 2D projections of non-cylindrically
symmetric distributions. This data compares well to the calculated
distributions, fig. \ref{fig:tomo-data}(b) \& (d), with $R_{ll}$
based only on fitting of the less detailed 2D projections, confirming
that the full angular structure is well-determined by the fitted matrix
elements. The fit results were further validated at the level
of the $\beta_{LM}$ parameters, extracted as a function of radius,
from the tomographic data (eqn. \ref{eq:IBlm}); this analysis will
be discussed in a future publication \cite{hockett2014}. It is anticipated
that, with the use of rich tomographic data-sets, the methodology
discussed here may be applied to more complex systems, where a greater
number of partial waves may play a role and 2D imaging data may not
prove sufficient for robust analysis.

\begin{figure}
\begin{centering}
\includegraphics[scale=0.7]{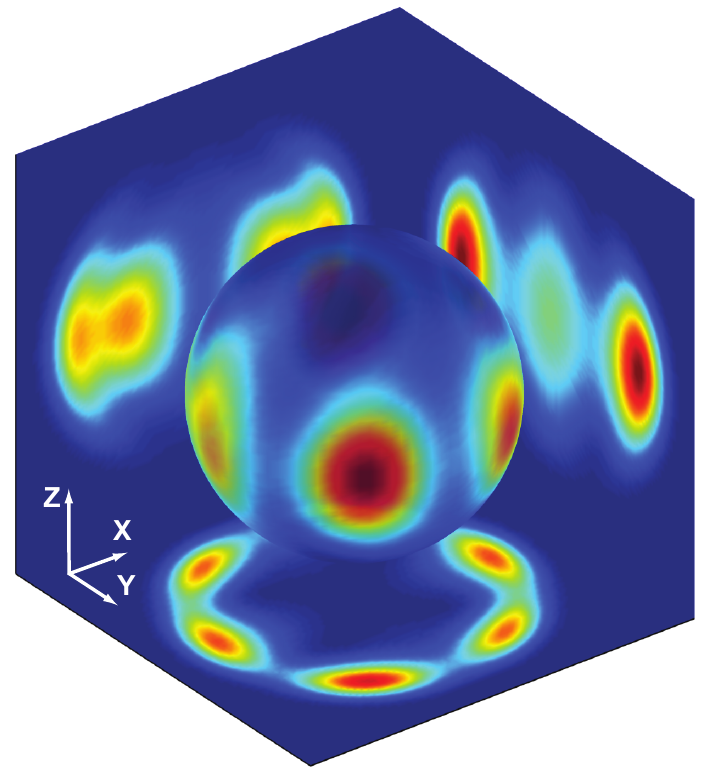}
\par\end{centering}

\caption{Application using a polarization shaped pulse. Calculated PAD
for $E_{X,Y}(t)$ obtained by application
of a $-\pi/2$ spectral phase mask to the red half of the spectrum, similar to the experimental pulses discussed in ref. \cite{Wollenhaupt2009a}. The full $I(\theta,\phi)$ distribution is shown in spherical projection, along with 2D projections onto different
Cartesian image planes.\label{fig:Application-to-polarization}}

\end{figure}

Finally, we illustrate the application of the ionization matrix elements
determined above to the case of a more complex polarization-shaped
pulse, shown in figure \ref{fig:Application-to-polarization}. 
For such polarization-shaped
pulses, the PAD can be considered as a sum over the ``basis states''
given by the PADs correlated with individual polarization states,
as defined by the $\beta_{LM}(\phi_{y})$ expansion in figure \ref{fig:images}(e);
an experiment utilizing a polarization-shaped pulse thus constitutes
a highly-multiplexed interrogation of the light-matter interaction in
polarization space. Consequently the PAD may be highly structured,
and is extremely sensitive to the exact shape of the laser pulse.
The 2D projections may show even greater sensitivity to the pulse
shape than the full $I(\theta,\phi)$ distribution, due to the enhanced
effect of rotations of the distribution on the image plane projections;
in this manner pulse shaping with imaging detection can lead to a
wide range of 2D projections, even in the case where only a few partial
wave channels are accessed in the ionization continuum.

In this work we have demonstrated the utility of complex intra-pulse
light-matter interactions as a means to ``complete'' photoionization
experiments, requiring only a few experimental measurements, combined
with theoretical treatment of the intra-pulse dynamics, to obtain
a set of ionization matrix elements. The validity of the radial matrix
elements determined from the experimental data were further tested
by comparison with tomographic (3D) data. In the case of potassium
multiple pathways play a role at the 1 and 2-photon level, with $p\rightarrow d\rightarrow f$
the major ionization path. With the ionization matrix elements to
hand calculations of the PAD for polarization-shaped laser pulses
become possible, and this was illustrated for a specific spectral
phase mask. This methodology is applicable to any arbitrarily shaped
laser field, providing the ionization can be treated perturbatively
and the photoelectron energy spread is small (with respect to the
response of $R(k)$), thus enabling a route to designing control
fields with full understanding of the control process. Combined with
tomographic reconstruction, or other means of obtaining full 3D PADs,
this should be a powerful technique for understanding the ionization
dynamics in many atomic systems, and should be extensible to molecular
ionization dynamics, including multiphoton ionization of chiral molecules (which exhibits photoelectron circular dichroism \cite{Lux2012,Janssen2014}) - provided that the population dynamics can be accurately modelled.

Acknowledgements: We thank Albert Stolow for suggesting this collaboration.

\bibliographystyle{apsrev4-1}
\bibliography{bibliography/baumert_collab_final_noURL.bib}

\end{document}